\documentclass[aps, reprint, longbibliography]{revtex4-2}
\usepackage{graphicx, natbib, amsmath}
\usepackage{subfigure, amssymb}
\usepackage{float} 

\DeclareGraphicsExtensions{.pdf,.eps,.png,.jpeg,.mps}
\graphicspath{ {Yilin-Figures/} }

\begin{document}

\title{Thermoelectric effects at a Germanium-electrolyte interface: \\
measuring $100 \, nK$ temperature oscillations at room temperature}
\author{Yilin Wong}
\author{Giovanni Zocchi}
\email{zocchi@physics.ucla.edu}
\affiliation{Department of Physics and Astronomy, University of California - Los Angeles}

\begin{abstract}
\noindent We describe measurements of $100 \, nK$ temperature oscillations at room temperature, 
driven at the complex interface between p-doped Germanium, a nm size metal layer, and an electrolyte. 
We show that heat is deposited at this interface by thermoelectric effects, however the precise microscopic 
mechanism remains to be established. The temperature measurement is accomplished by observing 
the modulation of black body radiation from the interface. We argue that this geometry offers a method 
to study molecular scale dissipation phenomena. The Debye layer on the electrolyte side of the interface 
controls much of the dynamics. Interpreting the measurements from first principles, we show that 
in this geometry the Debye layer behaves like a low frequency transmission line.
\end{abstract}

\keywords{Debye layer; Peltier effect; thermoelectric coefficients; THz spectroscopy}

\maketitle

{\bf Keywords} Debye layer, Peltier effect, thermoelectric coefficients, THz spectroscopy

\section  {Introduction} 

\noindent Boundary layers are ubiquitous non-equilibrium structures at solid-fluid interfaces. 
They are  in general loci of interesting dynamical behavior, through their transport properties and instabilities. 
The boundary layer is a region of large gradient of a field: in a fluid mechanics context this is typically 
the velocity field, but also the temperature field in the case of thermal convection. In solid state, we have 
the depletion layer at a semiconductor junction. In a neutral electrolyte, the Debye layer at a solid - fluid interface 
is similarly a region of large gradient of the electrostatic potential, i.e. large electric field. 
Here we describe two different dynamical effects which arise 
in an experimental setup where we force the Debye layer at low frequencies with an external 
electric field. In the experiment, the role of the Debye layer is essentially that of a distributed 
capacitance: it is the source for the capacitive current which drives the thermoelectric effects 
which we observe. By ``thermoelectric effects" we mean generally heat sources 
at an interface which are proportional to the electrical current, not the current squared. 
The Debye layer itself, under the circumstances of the experiment, behaves like an RC transmission line, 
which is the second somewhat unfamiliar dynamical effect which we describe. \\

\begin{figure}
	\centering
	\subfigure[Ge Chamber][\label{fig:GeChamber}]{\includegraphics[width=3in]{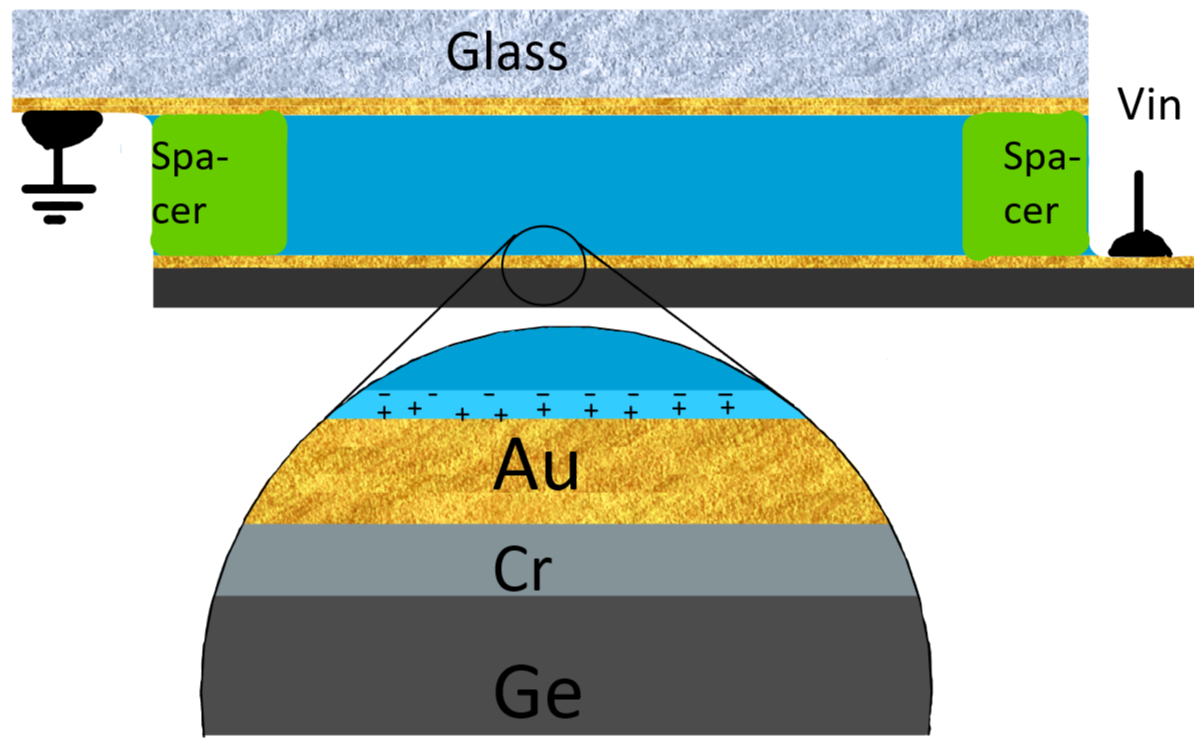}}
	\subfigure[Distributed RC][\label{fig:DistributeRC}]{\includegraphics[width=3in]{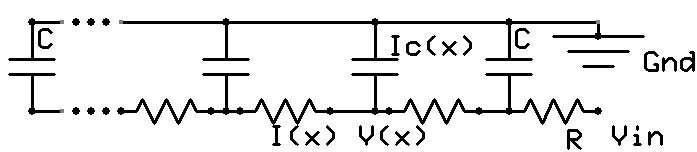}}
	\caption{(a) Sketch of the electrolyte-filled chamber (not to scale), constructed from a 
		$2 \times 5 \, cm$ piece of p-doped Ge wafer $170 \, nm$ thick, and a glass slide, separated 
		by $\sim 100 \, \mu m$ thick spacers. The inset shows the structure of the 
		Ge-metal-electrolyte interface, 
		consisting of a $3 \, nm$ layer of Cromium evaporated on the Germanium, and a $4 \, nm$ layer 
		of gold; also indicated is the $\sim 1 \, nm$ scale Debye layer (lighter blue) in the electrolyte. 
		The gold layer is, in reality, not uniform, consisting of ``islands" (see Mat. \& Met.). 
		The contact carrying the driving voltage is at one end of the chamber, as shown. 
		A thick ($30 \, nm$) gold layer on the glass slide forms the upper electrode. \\ 
		(b) The Ge-metal-Debye-layer strip is electrically similar to an RC transmission line: 
		the longitudinal resistance comes from the Ge and metal layers, the capacitance comes from 
		the Debye layer. The capacitive current $I_c$ is responsible for the thermoelectric effect. }
	
\end{figure}

Thermoelectricity has been known for two centuries, yet to this day there are new developments 
in the field (for a relatively recent review see \cite{Shakouri2011}), especially in the context of designing materials 
and geometries with optimized properties for energy conversion \cite{Hicks1993, Ohta2007, May2021}. Thermoelectric devices at the sub-$\mu m$ scale represent a new and interesting field of research, 
also leading to the desire to develop new techniques for local temperature measurements at the 
$10$ to $100 \, nm$ scale \cite{Regan2020}. 
Thermoelectric effects at the solid - liquid interface, more specifically, are relevant 
in a variety of different settings, 
from the technology of crystal growth \cite{Wiegel1997} to devices such as 
electrolyte gated transistors \cite{Palazzo2015, Dorfman2020}. \\
Our experiment was originally concieved to study dissipative phenomena at the solid - liquid 
interface, so we set up to measure far infrared (IR) radiation emitted from that region. 
Here we show that in our geometry, one can measure driven temperature oscillations of the solid-liquid interface 
of order $100 \, nK$ at room temperature. This resolution should allow to investigate by this method dissipative 
phenomena such as the internal dissipation of a driven molecular layer \cite{Amila2014, Zahra2018}, and in general hydrodynamic dissipative phenomena down to the molecular scale. \\

\section {Experimental results} 
\begin{figure}
	\centering
	\includegraphics[width=3in]{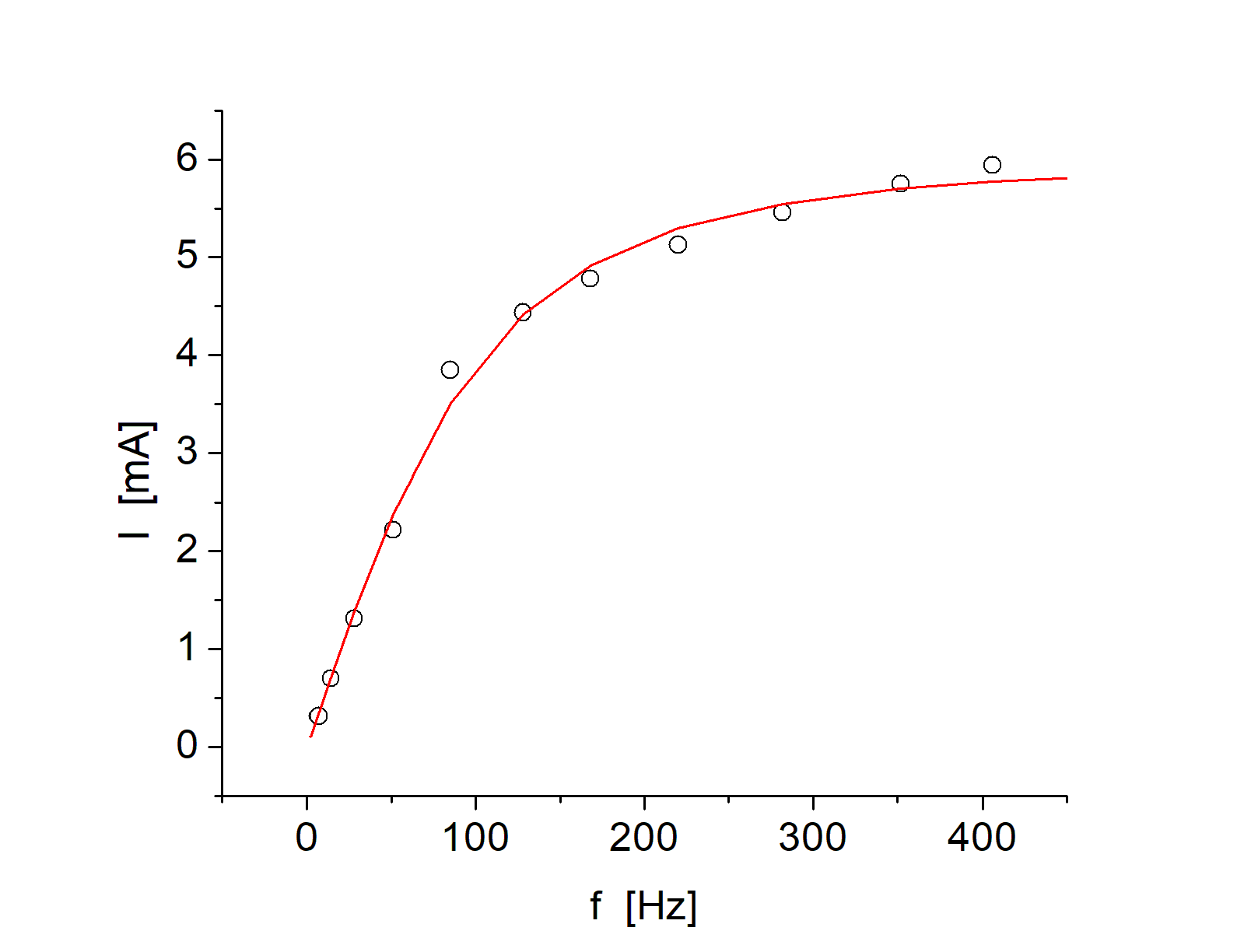}
	\caption{Current injected into the chamber vs driving frequency, for an amplitude of the drive of $0.5 \, V$. 
The solid line is a fit to an RC response, returning the values $R =84 \, \Omega $ , $C = 16 \, \mu F$ for 
the overall resistance and capacitance of the cell.}
	\label{fig:current}
\end{figure}

\noindent The setup we wish to consider consists of a thin conductive slab in contact with an aqueous 
electrolyte. Specifically, we start from a p-doped Ge wafer of thickness $\Delta = 170 \, \mu m$. Ge is 
used because (unlike water) it is relatively transparent to THz radiation in the wavelength range 
we wish to detect ($\lambda \sim 2 - 20 \, \mu m$). Using vacuum evaporation we deposit on one side 
of the wafer a $3 \, nm$ thick Cr layer followed by a $4 \, nm$ Au layer. The wafer is then cut into 
pieces roughly $5 \, cm \, \times \, 2 \, cm$ in size, and a rectangular 
chamber is constructed having the Ge chip as its ``bottom", a gold evaporated microscope slide as its 
``top", separated by $\sim 100 \, \mu m$ thick spacers (Fig. \ref{fig:GeChamber}). 
Electrical contacts are added at one end of this construction, 
and the chamber is filled with the electrolyte (a buffered solution of saline-sodium citrate (SSC) in water, 
see Mat. \& Met.). The Ge side of the chamber is placed directly in front of the entrance window of a 
liquid nitrogen cooled HgCdTe far infrared detector. The geometry is such that the $\sim 1 \, mm^2$ 
size detector ``sees" only a small area at the center of the chamber. The experiments consist of exciting 
the chamber with a sinusoidal voltage in the frequency range $1 \, Hz - 1 \, kHz$ (and amplitude 
$\sim 0.5 \, V$, below the threshold for water hydrolysis) and recording the intensity of emitted IR radiation 
in a phase locked loop. The measurement is sensitive to radiation in the wavelength 
range $2 - 20 \, \mu m$ ($15 - 150 \, THz$); for comparison, a room temperature photon has a wavelength 
$\sim 50 \, \mu m$. \\
Fig. \ref{fig:current} shows the amplitude of the current injected into the device vs the frequency of the applied voltage. Overall the response is that of an RC circuit (solid line), with $R = 84 \, \Omega$ , $C = 16 \, \mu F$ , 
$R C = 1.34 \times 10^{-3} \, s$. 

The resistance comes from the longitudinal resistance of the Ge and metal layers, while the capacitance 
comes from the Debye layer. The latter can be estimated as: $C = \epsilon S / (4 \pi \delta )$ (in esu), 
where $\epsilon \approx 80$ is the dielectric constant of water, $S \sim 1 \, cm^2$ the surface of the chamber, 
$\delta \approx 1 \, nm$ the Debye length; this gives $C \sim 100 \ \mu F$. The equivalent discrete elements 
circuit is an rc transmission line, sketched in Fig. \ref{fig:DistributeRC}. 
It is the capacitive current in this transmission line 
which drives the thermoelectric effect we will now describe. This capacitive current originates from switching 
the polarity of the Debye layer as the voltage alternates, and is carried by ions in the electrolyte, and by 
electrons and holes in the metal layers and the p-doped Germanium. \\
Our main result is displayed in Fig. 3, which shows the intensity of IR radiation seen by the detector vs 
frequency of the driving voltage. There are several notable features. First, this is the signal at the first harmonic 
(same frequency $\omega$ as the drive); there is no measurable signal from the IR detector at the second 
harmonic ($2 \omega$), or any other higher harmonics. Thus the IR emission does not come from Joule 
heating in the device (which would create a signal at the second harmonic), or, to say it differently, the 
infrared emission is proportional to the current, not the current squared. Second, what might look like a 
resonance is in fact the combination of two different phenomena: the increase in the capacitive current 
with frequency, and thermal diffusion away from the quasi-2D, solid - liquid interfacial region which is being 
alternately heated and cooled by thermoelectric effects. 

\begin{figure}
	\centering
	\includegraphics[width=3in]{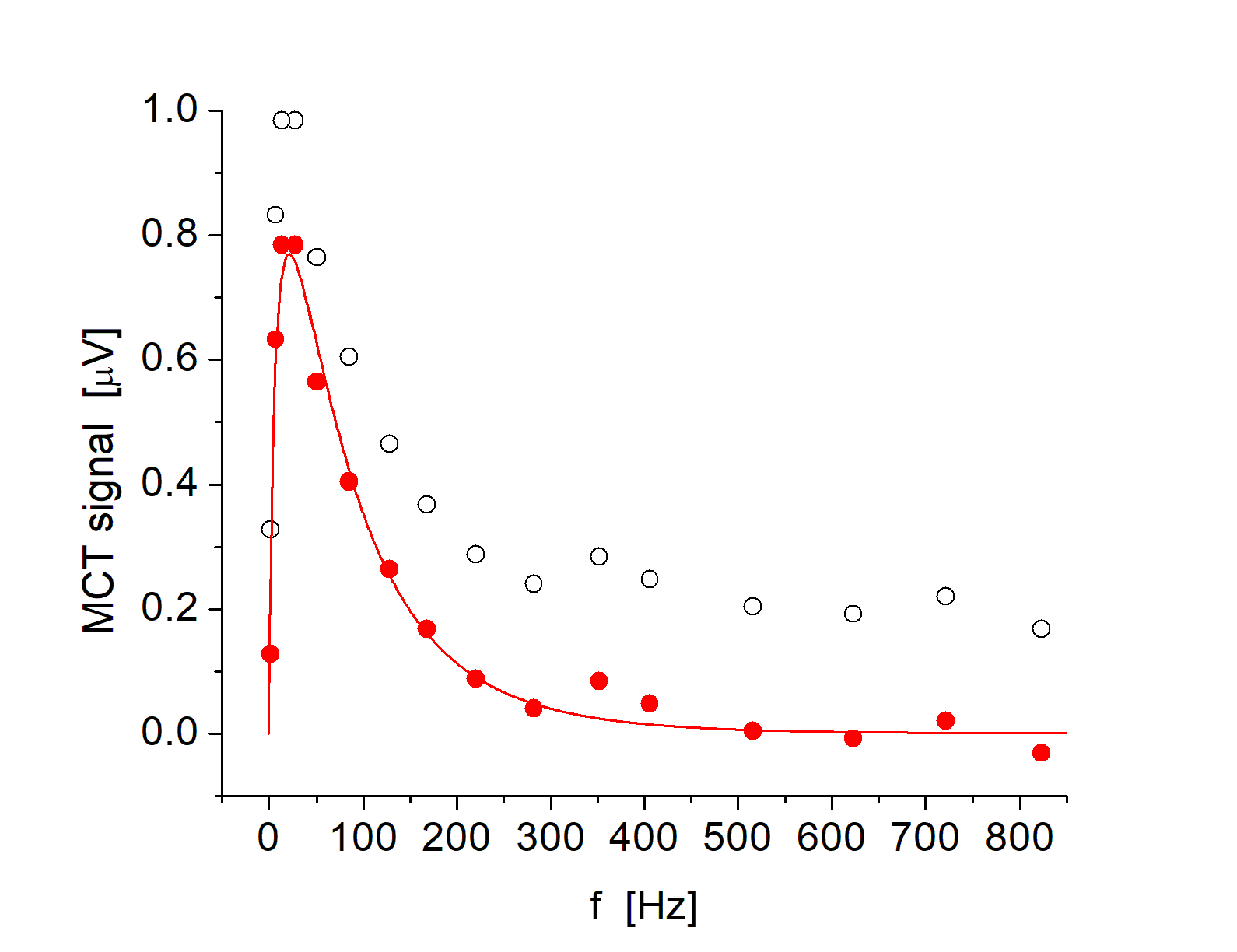}
	\caption{The infrared signal at the first harmonic vs driving frequency, for an amplitude of the drive of $0.5 \, V$. 
The current injected into the device is shown in Fig. \ref{fig:current}. The open circles represent the signal 
from the MCT detector after subtraction of a frequency dependent inductive background; the filled 
circles are the same data after subtraction of the plateau at high frequency. The solid line is a two-parameters 
fit with the theory developed in Section III. According to the calibration explained in the text (Section III), 
a $1 \, \mu V$ signal on this plot corresponds to an amplitude of temperature 
oscillations at the semiconductor - electrolyte interface of approximately $350 \, nK$ . }
	\label{fig:IT}
\end{figure}

The open circles in Fig. \ref{fig:IT} represent the measurements of intensity of IR radiation reaching the detector, 
after subtraction of a background which comes from inductive pickup, from the circuit driving the device, 
and from the device itself. In practice, we measure this background at various drive frequencies by blocking 
the entrance window to the detector; the measured background is proportional to the drive frequency 
$\omega$ : the corresponding straight line has been subtracted from the intensity measurements to obtain 
the open circles in Fig. \ref{fig:IT}. The filled circles are the same data as the open circles, 
after subtraction of a constant 
(independent of $\omega$) electronic offset determined by requiring $I(\omega) \rightarrow 0$ for 
$\omega \rightarrow \infty$ , that is, we subtract the value of the high frequency plateau seen in the data 
points represented as open circles. That plateau results from imperfect offset trimming of the operational amplifier 
in the detection circuit, and is also of the same order (twice as large) as the intrinsic noise of the experiment 
due to fluctuations in black body background radiation. \\
Based on the quantitative analysis of the next section, our interpretation of these measurements is that 
the capacitive current due to building up the Debye layer in the elctrolyte alternately heats and cools the 
quasi-2D region at the solid - liquid interface, through a thermoelectric effect. The amplitude of the corresponding 
temperature oscillations depends on the driving frequency $\omega$, because of heat diffusion away from 
the interface. The temperature oscillations produce a corresponding amplitude modulation of the black body 
radiation emitted from the region of the interface, which is the signal we detect.

\section{Theory} 
\noindent We consider a situation where the 2D solid - liquid interface is heated (or cooled) by a heat current 
$j_h$ proportional to the capacitive current $j_c$: 

\begin{equation}
\vec j_h = \chi \, \vec j_c
\label{eq:heat_current}
\end{equation} 

\noindent where $\chi$ is a thermoelectric coefficient (the Peltier coefficient if the mechanism is the Peltier effect). 
Choosing a z axis perpendicular to the interface (at $z = 0$) and going into the Germanium, the capacitive current 
on the Ge side ($z \ge 0$) is 

\begin{equation}
j_c = j_0 \ cos(\omega t) H(z)
\label{eq:drive_current}
\end{equation} 

\noindent in the z direction; $H(z)$ is the step function ($= 1$ for $z > 0$ and $= 0$ for $z \le 0$). 
Writing the conservation of energy then leads to the following diffusion equation for the temperature field 
(on the Ge side): 

\begin{equation}
\frac{\partial}{\partial t} T(z, t) - D \frac{\partial^2 T}{\partial z^2} = - \alpha \, cos(\omega t) \, \delta (z)
\label{eq:diffusion_1}
\end{equation} 

\noindent where $D$ is the heat diffusion constant of the Germanium, and 

\begin{equation}
\alpha = \frac{\chi}{\rho c_v} \, j_0
\label{eq:alpha_Peltier}
\end{equation} 

\noindent $\rho$ and $c_v$ are the density and specific heat for Ge, $\delta$ is the Dirac delta function. 
Writing $T(z, t) = \theta (z) e^{i \omega t}$ (with the convention that the physical quantities are the real 
part of the complex ones) the general solution of (\ref{eq:diffusion_1}) is: 

\begin{equation}
\theta (z) = \frac{- \alpha}{4 D} \, \frac{i-1}{k} e^{- \gamma z} + A e^{\gamma z} + B e^{- \gamma z}
\label{eq:Temp_1}
\end{equation}

\noindent where 

\begin{equation}
k = \sqrt{\frac{\omega}{2 D}} \quad , \quad \gamma = (1 + i) k 
\label{eq:Temp_1b}
\end{equation}

\noindent with the constants $A$ and $B$ to be determined by the boundary conditions. \\
Integrating (\ref{eq:diffusion_1}) in the neighborhood of $z = 0$ gives: 

\begin{equation}
\theta ' (0^+) - \theta ' (0^-) = \frac{\alpha}{D}
\label{eq:bc_1}
\end{equation}

\noindent In the experiments, the heat flow problem at $z = 0$ (the solid - liquid interface) and $z = \Delta$ 
($\Delta$ is the thickness of the Ge) is not well defined, because neither do we control the heat current nor 
the temperature. We will assume symmetric conditions at $z = 0$: $\theta ' (0^+) = - \theta ' (0^-)$ , then 
using (\ref{eq:bc_1}) the boundary condition at $z = 0$ is 

\begin{equation}
\theta ' (0) = - \frac{\alpha}{2 D}
\label{eq:bc_2}
\end{equation}

\noindent Alternatively, if we assume no heat flux into $z < 0$, then the boundary condition is 
$\theta ' (0) = - \alpha / D$. At the other end ($z = \Delta$), we assume efficient removal of extra heat 
by convection in the air, i.e. 

\begin{equation}
\theta (\Delta) = 0
\label{eq:bc_3}
\end{equation}

\noindent With these boundary conditions, the solution (\ref{eq:Temp_1}) becomes: 

\begin{equation}
\theta (z) = (1 - i) \, \frac{\alpha}{4 D} \, \frac{1}{k} \, 
\frac{e^{\gamma (\Delta - z)} - e^{- \gamma (\Delta - z)}}{e^{\gamma \Delta} + e^{- \gamma \Delta}}
\label{eq:Temp_2}
\end{equation} 

\noindent Thus the amplitude of the temperature oscillation at the interface is 

\begin{equation}
|\theta (z = 0)| = \frac{\sqrt{2} \alpha}{4 D} \, \frac{1}{k} \, 
\left | \frac{e^{\gamma \Delta} - e^{- \gamma \Delta}}{e^{\gamma \Delta} + e^{- \gamma \Delta}} \right |
\label{eq:Temp_interface}
\end{equation} 

\noindent We are also interested in the temperature oscillation averaged over the Ge thickness; the corresponding 
amplitude is, using (\ref{eq:Temp_2}) : 

\begin{equation}
| \left < \theta \right > | = \left | \, \frac{1}{\Delta} \, \int_0^{\Delta} \theta (z) dz \, \right | = 
\frac{\alpha}{4 D} \, \frac{\Delta}{(k \Delta)^2} 
\left | \frac{e^{\gamma \Delta} + e^{- \gamma \Delta} - 2}{e^{\gamma \Delta} + e^{- \gamma \Delta}} \right |
\label{eq:Temp_averaged}
\end{equation} 

\noindent We expect the signal detected in the experiments to be proportional to the expression 
(\ref{eq:Temp_averaged}) (or perhaps (\ref{eq:Temp_interface})), since the power emitted per unit surface 
from black body radiation, given a temperature modulation of amplitude $\theta$ around room temperature 
$T_0$, is: 

\begin{equation}
P = \epsilon \, \sigma \, ( T_0 + \theta )^4 \approx \epsilon \, \sigma \, T_0^4 \, 
\left (1 + 4 \frac{\theta}{T_0} \right ) = const. + 4 \epsilon \, \sigma \, T_0^3 \, \theta 
\label{eq:Black_Body}
\end{equation} 

\noindent $\sigma$ is the Stefan - Boltzmann constant and $\epsilon$ ($0 < \epsilon < 1$) the emissivity of the emitting surface. The parameter $\alpha$ in (\ref{eq:Temp_averaged}) and (\ref{eq:Temp_interface}) is 
proportional to the amplitude of the capacitive current (see (\ref{eq:alpha_Peltier})), which is given by 
the measured RC response of the cell (Fig. \ref{fig:current}): 

\begin{equation}
I_c = \frac{V_0}{R} \, \frac{\omega R C}{[1 + (\omega R C)^2]^{1/2}} \quad , \quad j_0 = \frac{I_c}{S}
\label{eq:RC_response}
\end{equation} 

\noindent where $S$ is the area of the cell. \\
A fit of the data in Fig. \ref{fig:IT} using either (\ref{eq:Temp_averaged}) or (\ref{eq:Temp_interface}) is 
a stringent requirement, since all parameters are known, except for an overall multiplicative constant 
(which contains the calibration of the IR detection system). However, upon attempting this one-parameter 
fit, it is apparent that the frequency dependence expressed by (\ref{eq:Temp_averaged}) or (\ref{eq:Temp_interface}) is not the correct one vis-a-vis the measurements. What is missing from our 
theory is consideration of how the driving voltage travels along the Debye layer, which introduces an 
additional frequency dependence to the capacitive current at the point of measurement. We now analyze 
this interesting phenomenon. Consider the Ge chip coupled to the electrolyte: the Ge (and metal 
layers) act as a distributed resistance, while the Debye layer acts essentially as a distributed 
capacitance. The result is an RC transmission line (Fig. \ref{fig:DistributeRC}), which we model 
in 1D; the governing ``cable equation" for the voltage $V(x, t)$ is then: 

\begin{equation}
\frac{\partial V(x, t)}{\partial t} - \frac{1}{r c} \, \frac{\partial^2 V}{\partial x^2} = 0
\label{eq:cable}
\end{equation} 

\noindent i.e. the diffusion equation with diffusion constant $1 / r c$ ; $r$ is the resistance per unit length, $c$ the capacitance 
per unit length. The capacitive current per unit length (i.e. the current going into the Debye layer) is 

\begin{equation}
I_c (x, t) = c \, \frac{\partial}{\partial t} V(x, t) 
\label{eq:capa_current}
\end{equation} 

\noindent We consider a sinusoidal input at $x = 0$: $V(x = 0, t) = V_0 \, exp [i \omega t]$ and, for simplicity, 
a semi-infinite cell (so that $V \rightarrow 0$ for $x \rightarrow + \infty$); then the solution of 
(\ref{eq:cable}) is: 

\begin{equation}
V (x, t) = V_0 \, e^{- \kappa x} \, e^{i (\omega t - \kappa x)}  \quad , \quad  \kappa = \sqrt{\frac{\omega r c}{2}}
\label{eq:voltage_wave}
\end{equation} 

\noindent Using (\ref{eq:capa_current}) and (\ref{eq:voltage_wave}), the capacitive current per unit length is: 

\begin{equation}
I_c (x, t) = i \omega c \, V_0 \, e^{- \kappa x} \, e^{i (\omega t - \kappa x)}  
\label{eq:capa_current_2}
\end{equation} 

\noindent so that 

\begin{equation}
| I_c (x) | =  \omega c \, V_0 \, e^{- \kappa L (x / L)} 
\label{eq:capa_current_2}
\end{equation}

\noindent where $x / L$ is the observation point with respect to the length of the cell $L$. To relate the 
physical system (a strip of length $L$ and width $w$) to this 1D model we note that the capacitive current 
density is $j_c (x) = I_c (x) / w$ while $c / w := c_A$ is the capacitance per unit area; $C = c L$ is the 
total capacitance of the cell, $R = r L$ the total resistance 
(these are measured in Fig. \ref{fig:current}). If we describe the thermoelectric 
effect through a Peltier coefficient $\chi$ (see (\ref{eq:heat_current})), then $\alpha = [\chi / (\rho c_v)] \, j_c$ 
and putting all the pieces together (\ref{eq:Temp_averaged}) reads: 

\begin{equation}
| \left < \theta \right > | = \frac{\chi c_A}{2 \rho c_v \Delta} \, V_0 \, 
exp \left [ - \sqrt{\frac{\omega R C}{2}} \, \frac{x}{L} \right ] \, 
\left | \frac{e^{\gamma \Delta} + e^{- \gamma \Delta} - 2}{e^{\gamma \Delta} + e^{- \gamma \Delta}} \right |
\label{eq:Temp_averaged_2}
\end{equation} 

\noindent $x / L$ is the observation point; we treat it as a geometric parameter of order 1, which we are 
allowed to fit (considering that the real system is 2D, with imperfectly known geometry of the contacts, etc.). \\ 
If instead of (\ref{eq:Temp_averaged}) we use (\ref{eq:Temp_interface}) we find the amplitude of the 
temperature oscillation at the interface as: 
\begin{equation}
|\theta (z = 0)| = \frac{\chi c_A}{2 \rho c_v} \,  \sqrt{\frac{\omega}{D}} \, V_0 \, 
exp \left [ - \sqrt{\frac{\omega R C}{2}} \, \frac{x}{L} \right ] \, 
\left | \frac{e^{\gamma \Delta} - e^{- \gamma \Delta}}{e^{\gamma \Delta} + e^{- \gamma \Delta}} \right |
\label{eq:Temp_interface_2}
\end{equation} 
Comparing (\ref{eq:Temp_averaged_2}) with the measurements of Fig. \ref{fig:IT} is a stringent test, as there 
are only two fitting parameters: an overall multiplicative factor (which encompasses the detector calibration, 
among other factors), and the dimensionless number $x / L$ (which is basically a geometric factor). All other 
parameters are measured or known: $\Delta = 170 \, \mu m$ (thickness of the Ge slab), 
$R C = 1.34 \times 10^{- 3} \, s$ (measured, see Fig. \ref{fig:current}), $D = 0.36 \, cm^2 / s$ 
(thermal diffusion constant for Ge). Fig. \ref{fig:IT} shows that the form (\ref{eq:Temp_averaged_2}) 
fits the measurements very well (solid line in the figure). Using the temperature oscillaton at the interface 
(\ref{eq:Temp_interface_2}), instead of the averaged quantity (\ref{eq:Temp_averaged_2}), results in an essentially 
identical fit. In summary, the mechanism we identified for the observed frequency dependence of the IR 
emission - namely, heat transport to and from the 2D solid-liquid interface by thermoelectric effects - appears 
to be correct. 

\section {Discussion} 
\noindent The fit of Fig. \ref{fig:IT} shows that the dependence of the emitted radiation on the driving frequency 
is well captured by the theory (\ref{eq:Temp_averaged_2}) or (\ref{eq:Temp_interface_2}), 
but it is also instructive to consider the 
absolute magnitude of the corresponding temperature oscillation. According to the factory calibration of our 
MCT detector (dark resistance $R_d = 55 \, \Omega$ at $77 \, K$), the change in resistance per unit 
incident intensity is $\delta R_d / P = 0.158 \ \Omega / (\mu W / cm^2)$ for black body radiation at 
$500 \, K$. With our electronics (see Mat. \& Met.) this figure translates to the following calibration for the 
measurements of Fig.  \ref{fig:IT}: \\ 
a signal $S = 1 \, \mu V$ corresponds to an incident intensity $P [S = 1 \, \mu V] = 54.4 \, pW / cm^2$ . \\
There is a correction because in the experiment the radiation is at $293 \, K$, which we ignore for the moment. 
Using the Stefan - Boltzmann law, the intensity (power per unit surface) of black body radiation emitted by 
the heated interface at the driving frequency $\omega$ is 
\begin{equation}
P = \epsilon \, 4 \sigma \, T_0^3 \, | \left < \theta \right > |
\label{eq:Stefan-Boltz}
\end{equation} 
where $T_0 = 293 \, K$, $\sigma$ is the Stefan - Boltzmann constant, and we are using the averaged amplitude 
of the temperature oscillation, $| \left < \theta \right > |$, for now. $\epsilon$ is the effective emissivity of 
the cell ($0 < \epsilon < 1$), which we discuss later. Thus the amplitude of temperature oscillation which, 
in the experiment, corresponds to a $1 \, \mu V$ signal is: 
\begin{equation}
| \left < \theta \right > | [S = 1 \, \mu V] = \frac{P [S = 1 \, \mu V]}{\epsilon \, 4 \sigma \, T_0^3} 
\label{eq:calibration_1}
\end{equation} 
If we assume for now $\epsilon = 1$ , this expression gives 
$| \left < \theta \right > | [S = 1 \, \mu V] \approx 95 \, nK$ . 
What is the actual effective emissivity to be used in (\ref{eq:calibration_1}) is not completely clear, because 
our cell is a composite Germanium - metal - water structure, but we reason as follows. The 
$\Delta = 170 \, \mu m$ thick Ge slab alone would have very small emissivity in the infrared range we are 
considering, because the slab is essentially transparent in that range. The inverse absorption length in Ge 
at a wavelength $\lambda \sim 10 \, \mu m$ is $\kappa \approx 0.02 \, cm^{-1}$ , which gives the fraction 
of intensity absorbed by the slab as $A = 1- e^{- \kappa \Delta} \approx 3 \times 10^{-4}$ ; this is also 
the emissivity (i.e. $\epsilon = A$). Ignoring the nm size metal layers, the water in close proximity to the interface 
is on the other hand an effective black body emitter. The emissivity of bulk water in the far infrared 
($\lambda \sim 10 \, \mu m$) is $\epsilon \approx 0.90$ , while the inverse absorption length is 
$\kappa \approx 10^3 \, cm^{-1}$ , thus already a $\ell = 1 / \kappa \approx 10 \, \mu m$ thick layer 
of water forms an effective black body emitter at our wavelengths. The thermal diffusion time across this 
water layer is $\tau \sim \ell^2 / D_w \approx 1 \, ms$ ($D_w \approx 1.4 \times 10^{-3} \, cm^2 / s$ is 
the thermal diffusion constant of water). Since the thermal diffusion constant of Germanium is more than 
100 times larger than $D_w$, the picture we arrive at is that while the temperature dynamics is controlled 
by the Ge slab, the black body emissivity is controlled by this $\sim 10 \, \mu m$ thick water layer at the 
interface, at least for driving frequencies $\nu \le 1 / \tau = 1 \, kHz$. Therefore we should use an emissivity 
$\sim 0.90$ , and taking into account a transmittance of the Ge exit surface  (the Ge - air interface) of 
$\sim 40 \, \%$ , we arrive at an effective emissivity for the cell of $\epsilon \approx 0.36$ . Also, from 
the preceding discussion we expect that what matters for the IR emission is the amplitude of the temperature 
oscillation at the interface, $| \theta (z=0) |$ , rather than the amplitude $| \left < \theta \right > |$
averaged over the Ge slab. With $\epsilon = 0.36$ we then find from (\ref{eq:calibration_1}) that 
the amplitude of temperature oscillation at the interface corresponding to a $1 \mu V$ signal in the experiment 
(see Fig. \ref{fig:IT}) is $ | \theta (z=0)| [S = 1 \, \mu V ; \epsilon = 0.36] \approx 264 \, nK $ . 
We should also take into account the difference between 
the factory calibration of the MCT detector using $500 \, K$ black body radiation 
and the experiment with radiation at $293 \, K$ (see Mat. \& Met.). From the spectral 
sensitivity of the detector given by the manufacturer and the black body spectra at $500 \, K$ 
and $293 \, K$ we find that the calibration above needs to be adjusted by a factor $1.34$ , 
so that the temperature oscillation corresponding to a $1 \, \mu V$ signal in the experiment is  
(assuming an effective emissivity $\epsilon = 0.36$) : \\
\begin{equation}
| \theta (z=0)| [S = 1 \, \mu V] \approx 354 \, nK . 
\label{eq:calibration_5}
\end{equation} 
With this calibration, the measured amplitude of temperature oscillation e.g. at $\nu = 50 \, Hz$ is, from 
Fig. \ref{fig:IT}, 
\begin{equation}
| \theta |_{measured} = 354 \, nK / \mu V \times 0.65 \, \mu V = 230 \, nK \quad at \quad \nu = 50 \, Hz
\label{eq:calibration_3}
\end{equation} 
Let us now see what is the magnitude of the temperature oscillation predicted by the theory 
(\ref{eq:Temp_interface_2}). Let us take again the driving frequency $\nu = 50 \, Hz$ 
$\Rightarrow \omega = 2 \pi \nu \approx 300 \, rad / s$. Since $D \approx 0.36 \, cm^2 / s$ , then 
$k = \sqrt{\omega / (2 D)} \approx 21 \, cm^{-1}$ (thus the wavelength of the damped voltage oscillation 
in the transmission line is $\lambda = 2 \pi / k \approx 3.1 \, mm$ at this frequency), and $k \Delta \approx 0.347$. 
Then $| \gamma \Delta | < 1$ (recall that $\gamma = (1 + i) k$) and expanding the exponentials one finds: 
 \begin{equation}
\left | \frac{e^{\gamma \Delta} - e^{- \gamma \Delta}}{e^{\gamma \Delta} + e^{- \gamma \Delta}} \right | 
\approx \sqrt{2} \, k \Delta
\label{eq:ampli_1}
\end{equation} 
and the expression (\ref{eq:Temp_interface_2}) becomes: 
 \begin{equation}
| \theta (z=0) | \approx \alpha \, \frac{\Delta}{2 D} \quad , \quad 
\alpha = \frac{\chi c_A}{\rho c_v} \, \omega \, V_0 \, 
exp \left [ - \sqrt{\frac{\omega R C}{2}} \, \frac{x}{L} \right ] 
\label{eq:ampli_2}
\end{equation} 
Incidentally, in the same approximation one finds that the expression for the averaged amplitude 
(\ref{eq:Temp_averaged_2}) differs from the present one only by an overall factor of 2: 
$| \theta (z=0) | = 2 \, | \left < \theta \right > |$. To evaluate $\alpha$, we use 
$c_A = (16 \, \mu F) / (4 \, cm^2) \approx 3.6 \times 10^6 \, cm^{-1}$ (in esu), from the measured 
capacitance  (Fig. \ref{fig:current}) and area of the cell. Further, $\rho c_v \approx 1.5 \times 10^7 \, 
ergs / (K cm^3)$ , $\Delta \approx 1.7 \times 10^{-2} \, cm$ , $V_0 = 0.5 \, V = 1 / 600 \, (esu)$ ; 
$x / L = 6.6$ (from the fit Fig. \ref{fig:IT}) , $R C = 1.34 \times 10^{-3} \, s$ (from the fit Fig. \ref{fig:current}), 
giving $exp \left [ - \sqrt{\frac{\omega R C}{2}} \, \frac{x}{L} \right ] \approx 4.83 \times 10^{-2}$ . 
The Seebeck coefficient for p-doped Ge, at our carrier concentration and room temperature, 
is approximately $700 \, \mu V / K$ \cite{Ohishi2016}. Through the Kelvin relation, the corresponding 
Peltier coefficient is $\chi \approx 200 \, mV$. Using this value, we obtain 
$\frac{\chi c_A}{\rho c_v} \, \omega \, V_0 \approx 8.4 \times 10^{-5}$ 
and finally $| \theta (z=0) | \approx 93 \, nK$ at $\nu = 50 \, Hz$ . 
This is a factor 2 smaller than the measured amplitude according to (\ref{eq:calibration_3}). 
However, given the lack of an independent experimental calibration of our temperature measurements, 
and also considering that the thermal boundary conditions leading to (\ref{eq:ampli_2}) are not well defined 
in the experiment, it is not clear that this discrepancy is significant. 
Nonetheless, we mention two other mechanisms which might contribute to the observed thermoelectric effect. 
The first is non-radiative electron-hole recombination in the depletion layer at the Ge - metal interface. 
Ge is an indirect gap semiconductor and non-radiative (phonon dependent) recombination 
dominates over direct transitions, especially in the presence of surface defects (``traps"). 
Junctions between p-doped Ge and a metal are typically ohmic, 
due to Fermi level pinning near the edge of the valence band \cite{Nishimura2007}. In such a junction, 
recombination in the depletion region can be the dominant conduction mechanism \cite{Hall1952, Rhoderick1982}. 
If electron-hole recombination is the origin of our thermoelectric effect, we simply should 
replace the Peltier coefficient $\chi$ in (\ref{eq:Temp_interface_2}) with 
$E / |e| \approx 670 \, meV$ where $E$ is the band gap, 
$|e|$ the elementary charge. Namely, the source term in the diffusion equation (\ref{eq:diffusion_1}) is, 
in the case of electron-hole recombination,
\begin{align*}
	S = (E / |e|) \,  j_c (t) \delta (z)
\end{align*}
using the notation of Section III; $j_c$ is the capacitive current. In the case of the Peltier effect, 
the source term is
\begin{align*}
	\vec \nabla \cdot \vec j_h = \chi \,  j_c (t) \delta (z)
\end{align*}
where $j_h$ is the heat current, so we see that $\chi$ turns into $(E / |e|)$  . With this modification in  
(\ref{eq:Temp_interface_2}) and (\ref{eq:ampli_2}) we now find: 
\begin{equation}
| \theta (z=0)| \approx 300 \, nK \quad at \quad \nu = 50 \, Hz .  
\label{eq:calibration_4}
\end{equation} 
This figure is now {\it larger} than the measurement (\ref{eq:calibration_3}) by about a factor of $2$ . 
But again, considering the uncertainties in estimating the effective emissivity of the cell, 
and also the somewhat arbitrary ``symmetric" boundary condition (\ref{eq:bc_2}) 
used in the calculation, there is reasonable agreement between theory and experiment on the 
absolute magnitude of the measured effect, whether we attribute (most of) this particular thermoelectric effect 
to electron-hole recombination or the the Peltier effect in the germanium.  \\
Yet another possibility is an anomalously large Peltier coefficient of the electrolyte, in our geometry. 
There have been recent theoretical proposals about large thermoelectric response of electrolytes 
at the scale of the Debye layer \cite{Dietzel2016, Fu2019}. These studies suggest that Seebeck coefficients 
as large as $\sim 5 \, mV / K$ may be obtained in electrolyte filled nanochannels \cite{Fu2019}, 
which corresponds to  
Peltier coefficients of order $\chi \sim 1.5 \, V$, up to 8 times larger than $\chi$ for our Germanium. 
This effect could thus be relevant to our measurements. While the situation considered in the above studies 
was for a flow of ions and corresponding temperature gradient parallel to the Debye layer 
(along the nanochannel), similar effects could arise in our geometry too. In general, the microscopic 
origin of the thermoelectric effect we measure is an interesting question, which can be answered 
experimentally, once we reopen our UCLA lab ``after the plague".\\
\noindent Let us summarize. We have introduced an experimental configuration where the Debye layer 
at the solid - electrolyte interface acts as the distributed capacitance of an RC transmission line, the 
semiconductor chip acting as the distributed resistive part. One motivation in constructing and analyzing 
such a system is that a similar configuration forms the passive part of the transmission line of  
the axon in the neuron \cite{Koch_Book}, where the cell membrane is the distributed capacitance, while 
the electrolyte in the confined space inside the axon acts as the distributed resistance. We have been 
experimenting with creating an artificial axon \cite{Hector2017, Hector2019}, and are thus interested 
in the dynamics of such systems. \\ 
An electrical current at the junction between two different conductors will in general give rise to 
thermoelectric effects. We were able to measure and quantitatively describe such a thermoelectric effect 
in a rather unfamiliar configuration. Consideration of similar effects is however relevant in the field of 
mixed electronic - ionic devices, for example, electrolyte gated transistors. In the course 
of this study we found that, in our simple setup, we are able to measure driven temperature fluctuations 
of the 2D solid-liquid interface of order $100 \, nK$ , at room temperature. There are not too many 
methods offering a combination of temperature resolution and space resolution in at least one dimension. 
For example, temperature mesurements with a space resolution of order $100 \, nm$ in all three dimensions 
are feasible using the optically detected electron spin resonance in nitrogen vacancy centers in diamond 
\cite{Neumann2013}; the temperature resolution is a fraction of $1 \, K$. 
Even higher spatial resolution in two out of three 
dimensions can be obtained  with electron microscopy based techniques, such as 
Plasmon Energy Expansion Thermometry \cite{Regan2015, Regan2017}, 
while the temperature resolution is still of order $1 \, K$. 
One can of course do much better in terms of temperature resolution if one abandons spatial resolution; 
for instance, a recent experiment based on a resonant optical cavity 
achieved a temperature resolution of order $1 \, nK / \sqrt{Hz}$ at room temperature \cite{Tan2017}. 
While breaking no barriers in either temperature or space resolution separately, the remarkable sensitivity 
of our setup in the context of a driven 2D interface should allow for studies of dissipative dynamics 
at the nm scale and room temperature. For instance, consider a layer of deformable 
macromolecules attached to the gold layer in our setup. As an example, take a globular protein of typical 
size $\sim 5 \, nm$. The protein can be deformed by the large electric field at the Debye layer 
\cite{Wang10}; deformations beyond the linear elasticity regime can be achieved \cite{WangPLoS, Qu12}, 
which are then dissipative \cite{Amila2014, Zahra2018}. If $n$ is the number of molecules per unit surface 
and $\epsilon$ the energy per molecule dissipated per cycle of the electric field, then the power dissipated 
per unit surface at the gold - electrolyte interface is $P = n \, \epsilon \, 2 \pi \omega$ . If we take as reasonable 
values $n = (1 \, molecule) / (5 \times 5 \, nm^2) = 4 \times 10^{12} \, cm^{-2}$ , 
$\epsilon = 4 \, kT = 0.1 \, eV$ (corresponding to breaking 1 - 2 hydrogen bonds while deforming the molecule), 
at $\nu = 50 \, Hz$ driving we get $P \approx 2 \times 10^{-2} \, mW / cm^2$ . By comparison, 
in the measurements of Fig. \ref{fig:IT} the power per unit surface deposited at the interface by the 
thermoelectric effects described is (if we consider electron-hole recombination as the mechanism) 
$S = (E / |e|) \, j_c$ where $j_c = V_0 \, c_A \, \omega \, 
exp \left [ - \sqrt{\frac{\omega R C}{2}} \, \frac{x}{L} \right ] $ is the capacitive current. At 
$\nu = 50 \, Hz$ driving, this gives $j_c \approx 3 \times 10^{- 2} \, mA / cm^2$ and 
$S \approx 2 \times 10^{- 2} \, mW / cm^2$ , same as the source term $P$ above. Thus, in this 
hypothetical example we would be able to detect a signal from the internal dissipation of a molecular layer 
with a signal over background ratio of order 1. In general, it seems possible to study dissipative phenomena 
at the scale of the Debye layer with this method.

\begin{acknowledgments}
\noindent This work was supported by NSF grant DMR - 1809381.
\end{acknowledgments}

\bibliography{Yilin_2_refs}

\end{document}


\title{Materials and Methods for: 
 ``Thermoelectric effects at a Germanium-electrolyte interface: \\
measuring $100 \, nK$ temperature oscillations at room temperature"}
\author{Yilin Wong}
\author{Giovanni Zocchi}
\email{zocchi@physics.ucla.edu}
\affiliation{Department of Physics and Astronomy, University of California - Los Angeles}

\maketitle

\section  {Deposition of Metal Layers} 
\noindent The $4 ''$, p-doped Ge wafer, of thickness $170 \, \mu m$, is cleaned with $O_2$ plasma for 10 min [Oxford Plasmalab]. The wafer is then immediately moved to an e-beam deposition machine [CHA Solution]. A layer of Cr is deposited at the rate of $0.2 \, \AA /s $ followed by a layer of Au deposited at 
the rate of $0.5 \, \AA / s$. The final nominal thicknesses are $3 \, nm$ for Cr and $4 \, nm$ for Au. 
These procedures were carried out at the UCLA Nanolab.\\
Glass microscope slides are sonicated for $15 \, min$ in acetone and for another $15 \, min$ in isopropyl alcohol. The slides are blow dried with nitrogen, then undergo the same procedure as the germanuim wafer. Namely, $10 \, min$ of $O_2$ plasma cleaning, immediately followed by e-Beam deposition. A layer of Cr and Au is deposited at the same rate as for the Ge wafer, but with final thickness of $3 nm$ and $30 nm$, respectively. In the experiments, this gold layer serves as the ground electrode. 

\begin{figure}[H]
	\centering
	\includegraphics[width=4in]{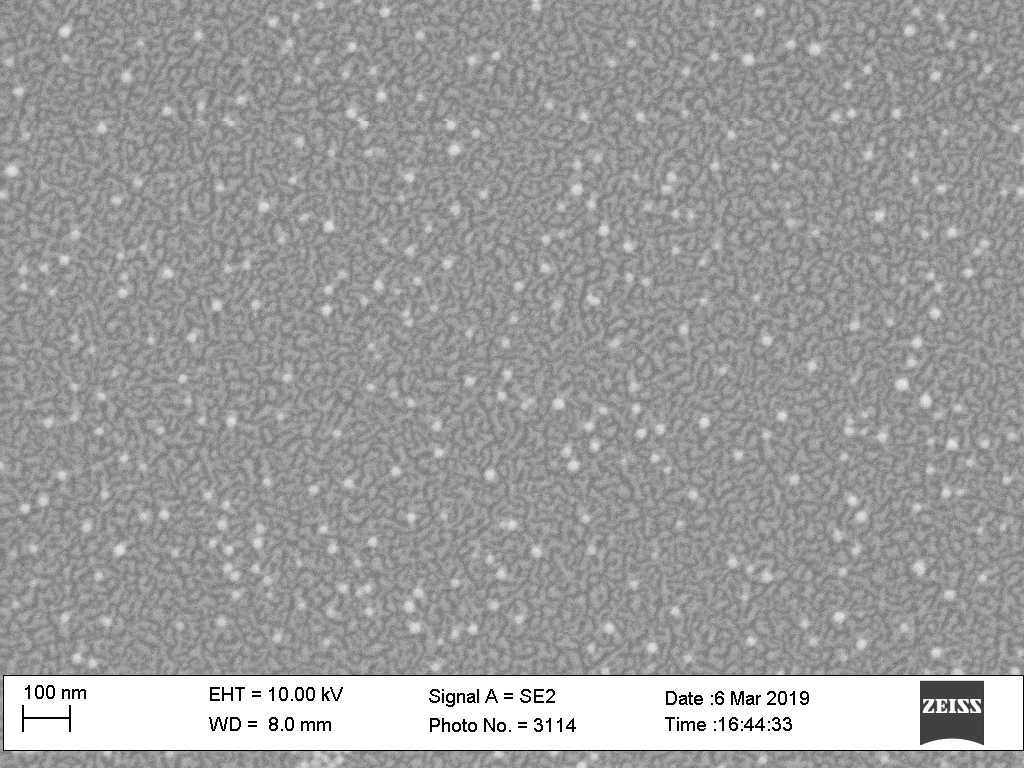}
	\caption{SEM picture of the Ge surface after deposition of the metal layers ($3 \, nm$ Cr and 
$4 \, nm$ Au). The light colored ``islands” are the gold, the darker background is the cromium. 
The white dots are $20 \, nm$ diameter gold particles used to facilitate focusing and size comparison.}
	\label{fig:SEM}
\end{figure}

Fig. \ref{fig:SEM} shows a SEM picture of the Ge surface after deposition. As expected, the Au layer 
is not uniform, consisting instead of ``islands" of typical size $10 - 50 \,nm$.
After deposition, the Ge wafer is cut into pieces roughly $5 \, cm \, \times \, 2 \,cm$ in size 
and stored in a clean container, while the slides are stored in $100 \, \%$ acetone.

\section {Chamber Construction} 
\noindent The first step in the construction of the chamber is to solder the electrical contacts 
on the metal layer of the Ge chip and the glass slide. We solder a copper wire onto one side of 
the Ge chip using low melting temperature Indium solder. This is a delicate operation, but it often  
(though not always) results in a good ohmic contact. 
Another copper wire is soldered onto one side of the microscope slide with regular solder. Both Ge and slide are then rinsed with acetone, isopropyl alcohol and finally DI water to remove any oil or dust that might be introduced in the previous process. 

\begin{figure}[H]
	\centering
	\includegraphics[width=4in]{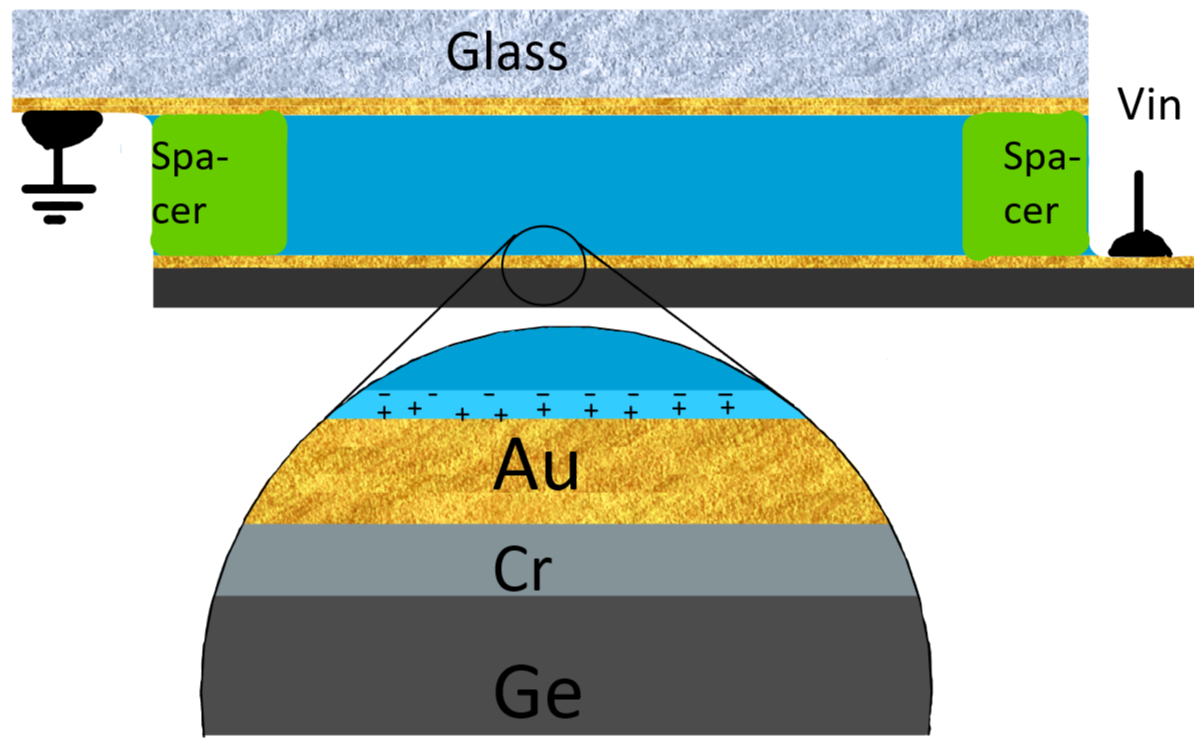}
	\caption{Rectangular Ge chamber with gold evaporated Ge chip on one side and microscope slide on the other side, seperated by $127 \mu m$ thick spacer.}
	\label{fig:Chamber}
\end{figure}

To form the chamber, the Ge chip and glass slide, separated by two $127 \, \mu m$ thick polyester spacers, are glued together using two-part epoxy, see Fig. \ref{fig:Chamber}. 
After the glue is dry, we use a pipette to fill the chamber with the electrolyte, carefully avoiding introducing air bubbles into the chamber. The electrolyte is SSC buffer (containing $150 \, mM$ sodium chloride 
and $15 \, mM$ trisodium citrate). After wiping off any access fluid, the remaining two ends 
of the chamber are sealed off with more two-part epoxy. The sealing procedure is repeated 2 to 3 times 
after each layer is dry to prevent leaking. 

\section{Geometrical/Electrical Setup}
\begin{figure}[H]
	\centering
	\includegraphics[width=4in]{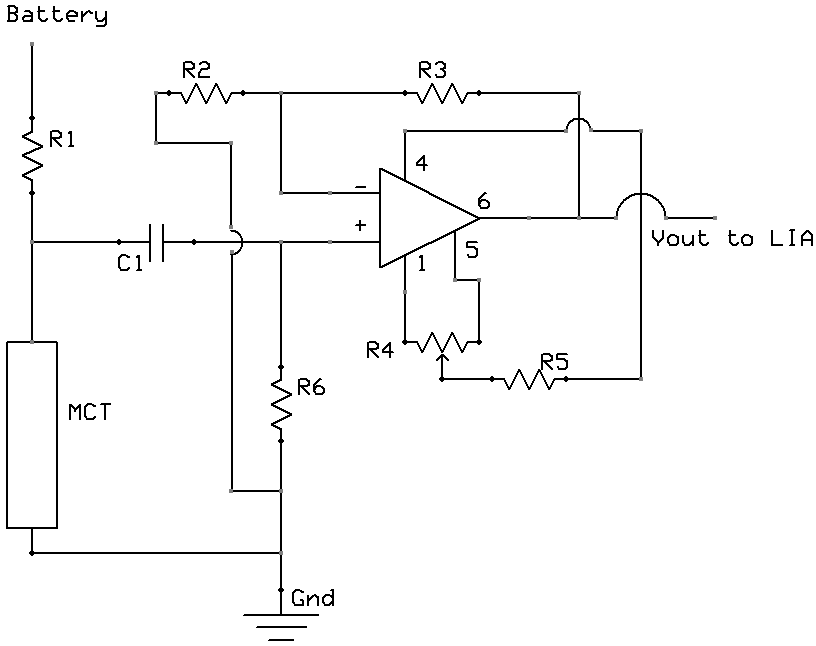}
	\caption{Detector circuit. $R1=1 \, k \Omega$, $C1=0.22 \, \mu F$, $R2=36 \, k \Omega$, 
$R3=330 \, k \Omega$, $R4=100 \, k \Omega$, $R5=1.5 \, k \Omega$, $R6=73 \, k \Omega$. 
MCT detector has a dark resistance of $55 \, \Omega$ when cooled to $70 \, K$. 
The OpAmp is the TL071. }
	\label{fig:MCT}
\end{figure}

\noindent Ge-Electrolyte chamber is positioned directly in front of the entrance window of the MCT detector 
(with the Ge side facing the detector),  with only a thin $\sim 2 \, mm$ air gap between the two. 
The current to the MCT detector is provided by a 12 V DC battery through a basic operation circuit as recommended by the manufacturer (Hamamatsu technical notes). As shown in Fig \ref{fig:MCT}, 
the detector is capacitively coupled to the sensing circuit, and trimming of the OP AMP offset is 
necessary.  The output of the OP AMP goes to a Lock-in amplifier. 

\begin{figure}[H]
	\centering
	\includegraphics[width=4in]{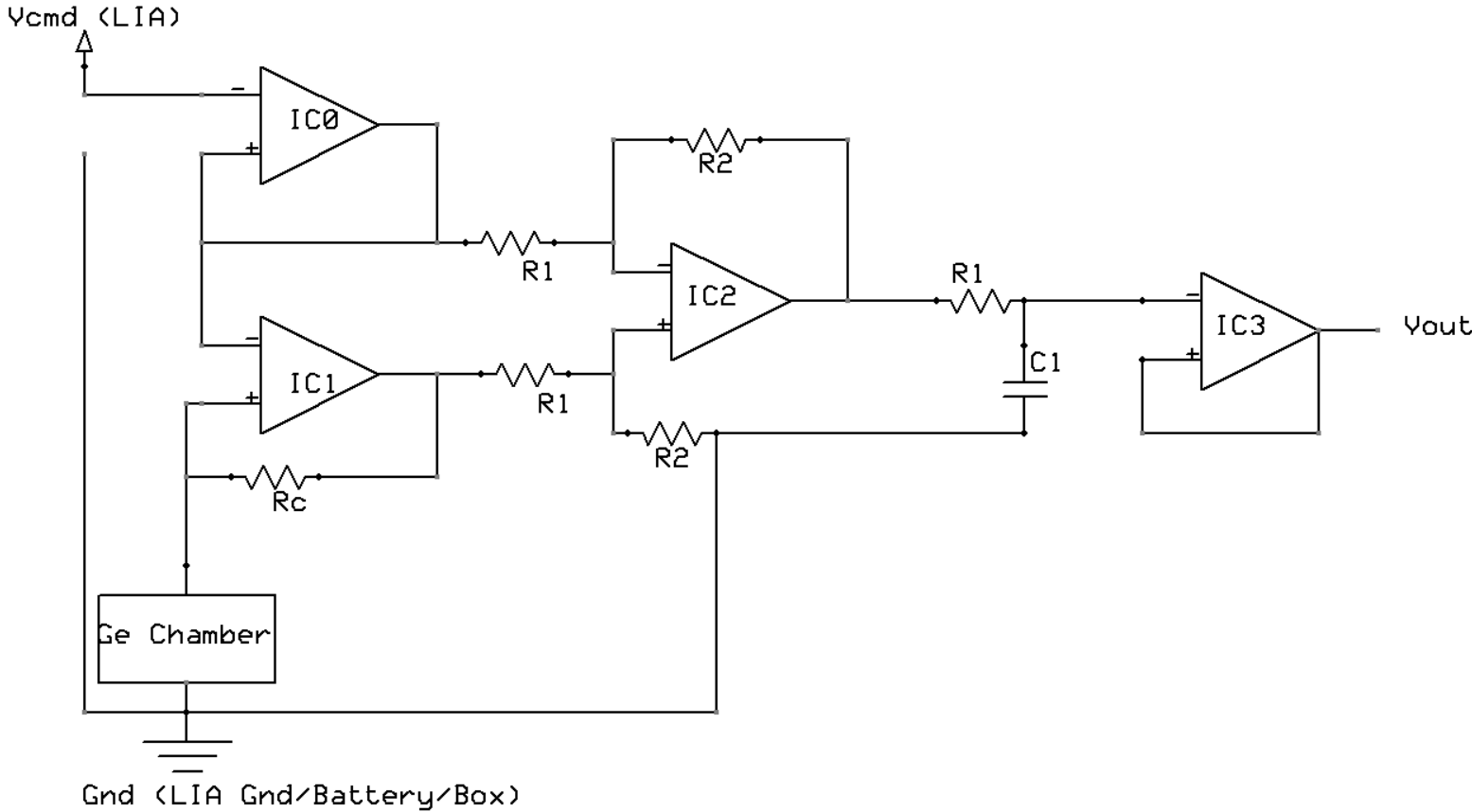}
	\caption{Schematics of the voltage clamp used to drive the chamber. The command signal 
$V_{cmd}$ is provided by the reference output of the Lock-in amplifier. The output $V_{out}$ 
is proportional to the current injected into the Ge chamber. 
$R1=1.2 \, k \Omega$, $R2=2.2 \, k \Omega$, $R_c=47 \, \Omega$, $C1=0.1 \, \mu F$. 
All OpAmps are TL071. }
	\label{fig:VClamp}
\end{figure}

To drive the cell and measure the current injected into the device we use a voltage clamp: the schematic is shown in Fig. \ref{fig:VClamp}. The output of this circuit is proportional to the current injected into 
the Ge chamber; the command voltage to the clamp is provided by the reference signal of the Lock-in 
amplifier. We drive the cell with a sinusoidal voltage, and record the current injected using  
an analogue to digital converter card [NI PCIe-6323].The recorded sinusoidal current is then used 
to calculate the rms current injected into the chamber. 

\section {H\MakeLowercase{g}C\MakeLowercase{d}T\MakeLowercase{e} Infrared Detector} 
\begin{figure}[H]
	\centering
	\includegraphics[width=4in]{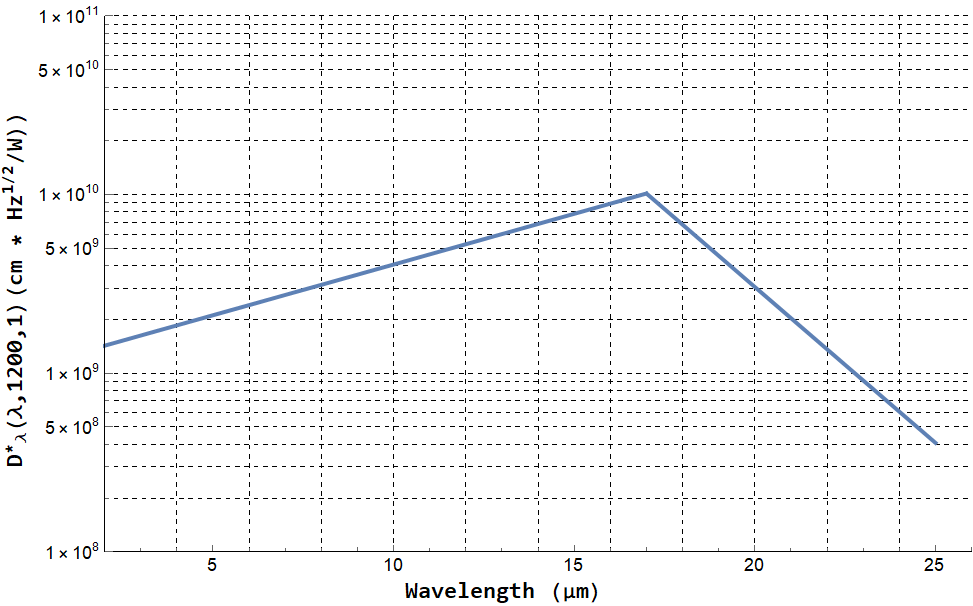}
	\caption{Spectral sensitivity of the MCT infrared detector, reproduced from the Hamamatsu 
technical notes. $D^*$ is the inverse of the noise equivalent power (NEP) in a $1 \, Hz$ bandwidth, 
multiplied by the square root of the detector area, in order to obtain a measure which is roughly 
independent of detector area.}
	\label{fig:Dstar}
\end{figure}

\noindent The detector we used is a quantum type photoconductive Mercury-Cadmium-Tellurium 
(MCT) detector [Hamamatsu P5274-01]. The band gap in this alloy material can be fine-tuned to 
give excellent sensitivity in the far infrared. 
In general, quantum type detectors have a wavelength-dependent sensitivity; 
the detector's resistance decreases when exposed to light due to the increase in carrier concentration.  
Since our particular model has been discontinued by Hamamatsu, we report in Fig. \ref{fig:Dstar}, 
for reference, the spectral sensitivity (the parameter D* in the literature) of the detector we use. 
According to the factory calibration, our detector, when tested with a 500 K black body source 
and incident intensity of $2.64 \times 10^{-6} \, W/cm^{2}$, yields a $2.5 \, mV$ signal 
for a detector current of $6 \, mA$. 
Considering our experiment is done at $293 \, K$, this calibration should be adjusted to account for the shift in peak radiant wavelength. D* is multiplied by blackbody spectral emittance and integrated 
from $0.6 \, \mu m$ to $25 \, \mu m$ to calculate the response of the detector to blackbody radiation at the corresponding temperature. Normalizing this response by the integrated blackbody emission we get 
the detector response per watt of incident radiation. Comparing this response at the two temperatures, 
we arrive at a $293 \, K / 500 \, K$ calibration factor of 1.34.

\section{Signal Processing}
\noindent We drive the chamber with a sinusoidal voltage at frequency $\nu$, thus black body emission 
from thermoelectric effects is also modulated at frequency $\nu$, and can be measured in a phase locked 
loop using a lock-in amplifier. For the measurements reported, we used a $3 \, s$ time constant 
for the output filter of the lock-in, and read out the signal on a digital oscilloscope, averaging for 
$\sim 1 \, min$.

\section{Other Specification of Ge Wafer}
\noindent Ga doped\qquad Orientation (100)\\
Carrier Concentration: $(1.88-4.13)\times 10^{16}/cc$\\
Mobility: $1520-1930 \quad cm^{2}/v.s.$\\
Polish: Double-side polished

\bibliography{Yilin_2_refs}